\documentclass[12pt]{article}

\usepackage{amsmath}
\usepackage{amssymb}
\usepackage{cite}

\usepackage{indentfirst}
\usepackage{amsfonts}
\usepackage{amscd}
\usepackage{amsbsy}
\usepackage{amsthm}
\usepackage{latexsym}
\usepackage{graphicx,color} 
\usepackage[dvipsnames]{xcolor}
\usepackage[colorlinks]{hyperref}
\hypersetup{linkcolor=blue,citecolor=blue,urlcolor=blue}

\makeatletter
\@addtoreset{equation}{section}

\makeatother

\newcommand{\nn}{\nonumber\\}
\def\beq{\begin{eqnarray}}
\def\eeq{\end{eqnarray}}
\newcommand{\bea}{\begin{eqnarray}}
\newcommand{\eea}{\end{eqnarray}}
\newcommand{\hs}[1]{\hspace{#1 mm}}

\def\ln{\,\mbox{ln}\,}

\def\det{\,\mbox{det}\,}
\def\tr{\,\mbox{tr}\,}

\def\Tr{\,\mbox{Tr}\,}

\def\al{\alpha}
\def\be{\beta}

\def\ga{\gamma}
\def\de{\delta}

\def\ep{\epsilon}

\def\la{\lambda}
\def\na{\nabla}

\def\si{\sigma}
\def\om{\omega}

\def\Ga{\Gamma}
\def\De{\Delta}

\renewcommand{\a}{\alpha}
\renewcommand{\b}{\beta}
\renewcommand{\c}{\gamma}
\renewcommand{\d}{\delta}
\renewcommand{\r}{\rho}
\newcommand{\s}{\sigma}
\newcommand{\G}{\Gamma}
\newcommand{\p}[1]{(\ref{#1})}
\newcommand{\RA}{\Rightarrow}
\newcommand{\pare}[1]{\left( #1 \right)}

\newcommand{\colc}[1]{\left[ #1 \right]}

\begin{document}

\begin{center}
\renewcommand{\thefootnote}{\fnsymbol{footnote}}

{\Large 
On the local term in the anomaly-induced action of Weyl
quantum gravity}
\vskip 4mm

{Andrei O. Barvinsky},$^{(a,b)}$\footnote{Email: \ barvin@td.lpi.ru }
\ \
{Guilherme H. S. Camargo},$^{(c,d)}$\footnote{Email: \ guilhermehenrique@unifei.edu.br}
\ \
{Alexei E. Kalugin},$^{(e)}$\footnote{Email: \ kalugin.ae@phystech.edu}
\ \
{Nobuyoshi Ohta},$^{(f,g)}$\footnote{Email: ohtan@ncu.edu.tw}
\ \
{Ilya L. Shapiro} $^{(c)}$\footnote{Email: \ ilyashapiro2003@ufjf.br}
\vskip 8mm

$^{(a)}$
Theory Department, P.N. Lebedev Physical Institute of the
Russian Academy of Sciences, Leninskiy Prospekt 53,  119991,
Moscow, Russia
\vskip 2mm

$^{(b)}$
Institute for Theoretical and Mathematical Physics,
Moscow State University, 119991, Leninskie Gory, GSP-1, Moscow, Russia
\vskip 2mm

$^{(c)}$
Departamento de F\'{\i}sica, ICE, Universidade
Federal de Juiz de Fora
\\
Juiz de Fora, 36036-330, MG, Brazil
\vskip 2mm

$^{(d)}$
Instituto de F\'{i}sica e Qu\'{i}mica, Universidade
Federal de Itajub\'{a}
\\
Itajub\'{a}, 37500-903, MG, Brazil
\vskip 2mm

$^{(e)}$
Moscow Institute of Physics and Technology, 141700,
Institutskiy pereulok, 9, Dolgoprudny, Russia
\vskip 2mm

$^{(f)}$
Department of Physics, National Central University, Zhongli,
Taoyuan 320317, Taiwan
\vskip 2mm

$^{(g)}$
Research Institute for Science and Technology, Kindai University,
Higashi-Osaka, Osaka 577-8502, Japan

\end{center}
\vskip 3mm


\begin{abstract}
\noindent
The finite local conformally non-invariant $R^2$-term emerges in the one-loop effective action of the model
of quantum gravity based on the Weyl-squared classical action. This term is related to the $\Box R$
contribution to the conformal anomaly, which in a wide class of regularization schemes is determined
by the second Schwinger-DeWitt (or Gilkey-Seeley) coefficient of the heat kernel expansion for inverse
propagators of the theory. The calculation of this term requires evaluating the contributions of
the fourth-order derivative minimal and of the second-order nonminimal operators in the tensor and
vector sectors of the theory, corresponding to metric, ghost and gauge-fixing operators.
To ensure the correctness of existing formulas, we derived (and confirmed) the result using
a special technique of calculations, based on the heat-kernel representation of the Euclidean
Green's function and the method of universal functional traces.
\vskip 2mm


\end{abstract}


\renewcommand{\thefootnote}{\arabic{footnote}}
\setcounter{footnote}{0}

\section{Introduction}
\label{sec1}

Conformal models play a very special role in modern quantum
field theory and there is an extensive literature about various aspects
of these theories. One of the important facts in even spacetime
dimensions is that the local conformal symmetry is violated by the
conformal (trace) anomaly $\langle T^\mu_{\,\mu} \rangle$, starting
from the one-loop level~\cite{CapDuf-74,duff77}.
Violation of local conformal symmetry comes in the form of local and nonlocal terms
in the effective action, generating this anomaly.
Breakdown of local Weyl invariance in classically conformally invariant theories with
$\langle T^\mu_{\,\mu} \rangle=0$ is the result of regularization and subtraction of UV divergences
by local diffeomorphism invariant counterterms. In curved spacetime, when gravity plays the role of
an external classical background, one-loop divergences of the classically conformal matter field theory has been
proven to be universally Weyl invariant~\cite{tmf}.  For Weyl-squared quantized conformal gravity,
the same statement has a more involved status. Initially it has been derived with the use of
the Fradkin-Vilkovisky conformization procedure~\cite{Fradkin-Vilkovisky} in~\cite{frts82},
then confirmed in~\cite{antmot} and passed verification by direct calculations in~\cite{Weyl}.\footnote{
Modulo quantum anomalies, local gauge invariance of counterterms generically follows to all loops
of perturbation expansion in the class of
{\em local background covariant} gauges~\cite{BRST}, but
their application in the case of Weyl squared quantum gravity
stumbles upon the problem of the search for such gauges satisfying
the condition of background covariance for both diffeomorphism
and {\em conformal} gauge transformations. Various approaches
to this problem include~\cite{frts82}, but can be circumvented
by direct calculations confirming Weyl invariance of one-loop
divergences in the conformally non-covariant background
gauges as was done in~\cite{antmot,Weyl,OP2015}. Gauge
independence of this result is discussed below in Sec.~3.}

Despite conformal invariance of one-loop divergences, their subtraction from the regularized effective action
entails nonvanishing $\langle T^\mu_{\,\mu} \rangle$, the structure of this trace anomaly reflecting
the structure of divergences. According to the general classification of the possible terms composing anomaly
in the purely metric sector of the four dimensional theory (we restrict ourselves with this dimension
only)~\cite{ddi,DeserSchwimmer}, the expression for $\langle T^\mu_{\,\mu} \rangle$ consists of the following
three types of terms:
\ {i)} conformally covariant square of the Weyl tensor
$C_{\mu\nu\al\be}^2=R_{\mu\nu\al\be}^2-2R_{\mu\nu}^2+\frac13R^2$ (its densitized version
$\sqrt{g}C_{\mu\nu\al\be}^2$ being conformally invariant);
\ {ii)} topological invariant Gauss-Bonnet term
$E_4=R_{\mu\nu\al\be}^2-4R_{\mu\nu}^2+R^2$ (its densitized version $\sqrt{g}E_4$ being conformally
invariant up to the addition of the total derivative term) and, finally,
\ {iii)} the total derivative $\Box R$-term.

All these structures, when they are densitized, i. e. multiplied by $\sqrt g$, represent the trace of
the functional variational derivatives of certain metric functionals. The principal difference between
them is that this functional is nonlocal for $\sqrt{g}C_{\mu\nu\al\be}^2$ and $\sqrt{g}E_4$ and local
only for the last total derivative contribution due to the relation
\beq
- \frac{2}{\sqrt{g}}\,g_{\mu\nu}
\frac{\delta }{\delta g_{\mu\nu}}\,\int d^4x\sqrt{g}\,R^2
\,=\, 12\,{\Box} R\,.
\label{identity-a}
\eeq
This relation shows that the $\Box R$-term in the quantum trace anomaly
can be modified or even completely removed by adding a finite term
$\int d^4x\sqrt{g}R^2$ to the {\it classical} action of gravity theory.
This is legitimate for quantum theory of conformal matter in external gravitational field,
where such a finite term belongs to the so-called vacuum part of the action (involving only
classical background metric field).
The vacuum action may not follow the symmetry of the quantum
fields and does not affect the number of active degrees of freedom.
Thus, the $\Box R$-term in the semiclassical anomaly is renormalization ambiguous (see e.g.~\cite{Duff94},
the detailed analysis of this issue in~\cite{anomaly-2004} and further developments for metric-scalar
models in~\cite{Asorey-2006}) and this opens the way for fixing
the coefficient of the $R^2$-term according to
observations as is usually done in high-energy particle physics,
e.g., related to the Starobinsky model of inflation~\cite{star,star83}
(see~\cite{anju,hhr,asta,BKN2015,StabInstab} for more recent
developments concerning the anomaly-induced inflation and
further references).

For quantized metric of Weyl invariant gravity theory, the situation is qualitatively different.
Its classical action is
\beq
S_W
&=&
-\,\int d^4x\sqrt{g}\,\Big\{\,\frac{1}{2\la}\,C^2
+ \frac{1}{2\rho}\,E_4+\frac1{2\xi}\,\Box R\,\Big\},
\label{action conf}
\eeq
where $C^2=C_{\mu\nu\al\be}^2$ is the square of the Weyl tensor,
$E_4=R_{\mu\nu\alpha\beta}^2-4R_{\mu\nu}^2+R^2$ is the Gauss-Bonnet integrand and $\la$, $\rho$ and $\xi$
are coupling constants of the dynamical, topological and total derivative terms. The action satisfies the
conformal Noether identity
\beq
-\,T^\mu_{\,\,\mu}
\,=\,  \frac{2}{\sqrt{g}}\,g_{\mu\nu}
\frac{\delta S_W}{\delta g_{\mu\nu}}\,=\,0.
\label{NoetherW}
\eeq
According to the existing tradition, $T^\mu_{\,\,\mu}$ is called
the trace of the metric stress tensor. Correspondingly,
the violation of the identity (\ref{NoetherW}) with $S_W$ replaced by the quantum effective action $\Gamma$
is called the quantum trace anomaly.

The different status of the theory (\ref{action conf}) from that of a conformal matter in external
gravitational field is that the counterterms needed to cancel the $C_{\mu\nu\al\be}^2$ and $E_4$ anomalies
are nonlocal, and their nonlocality contradicts the concept and the rules of renormalization by local
counterterms~\cite{Weinberg1960}. Now these metric functionals no longer belong to the vacuum (external field)
sector of the theory and carry quantum degrees of freedom of the theory. In higher-order loops of semiclassical
expansion the nonvanishing one-loop anomaly will irrecoverably destroy renormalizability of the theory and
its Ward identities providing its unitarity, as this was originally stated in~\cite{Capper1975,FrTs84}.
The $\Box R$ part of the anomaly will also make the theory inconsistent because the finite local
$R^2$-counterterm needed for its cancellation is itself conformally not invariant. Therefore, consistency
of the renormalization scheme would require introducing this term already at the classical level,
which would mean the loss of local Weyl invariance from the very beginning of the quantization procedure.

Despite inconsistency of the Weyl theory (\ref{action conf}) at the quantum level, there was much interest
in this model considered in the series of papers~\cite{frts82,antmot,Weyl,OP2015,OP2013}
where the Weyl squared and Gauss-Bonnet terms of the trace anomaly were fully calculated while
its $\Box R$ part was ignored. Lack of interest in this contribution might be explained by the fact that
it is usually considered to be ambiguous and depending on the chosen regularization and renormalization
scheme. For instance, in zeta function regularization~\cite{haw} or in the covariant cutoff of the lower
limit of the proper time integral~\cite{anomaly-2004}, the $\Box R$-term enters the anomaly as the trace
of the coincidence limit of the coefficient $\hat{a}_2$ of the Schwinger-DeWitt expansion, i.e., it is proportional
to the corresponding term in the one-loop divergence. On the contrary, in the dimensional regularization,
it is ambiguous and strongly depends on the details of analytic continuation into the complex plane of
spacetime dimensionality~\cite{duff77,birdav,Duff94}.

On the other hand, the $\Box R$ might be important in various implications because its contribution
to the finite nonzero $\langle T^\mu_{\,\mu} \rangle$ in view of Eq.(\ref{identity-a}) is responsible
for a finite $R^2$-term in effective action, and this term represents the core of the Starobinsky model
of inflation~\cite{star,star83}. In addition to this, the knowledge of $\Box R$ in the anomaly allows one
to pose the question of complete calculation of the surface terms in the one-loop divergences of the theory
and their dependence on the boundary conditions at spacetime boundaries. As the trace anomaly in fact represents
the spacetime integrand of the one-loop divergences --- the local Schwinger-DeWitt coefficient
$a_2$ (or Gilkey-Seeley $E_4$ coefficient)~\cite{DeWitt65,Gilkey}, its $\Box R$ part after integration
gets washed out from the spacetime bulk $\cal M$ and reduces to the surface integral
$\int_{\partial\cal{M}}d\Sigma^\mu\,\nabla_\mu R$ over the boundary $\Sigma=\partial{\cal M}$,
whereas another part of this surface integral comes from the boundary part of the Gilkey-Seeley
coefficient $E_3$. This issue was a subject of a very preliminary analyses~\cite{frts82} and
never fully considered within Weyl invariant gravity theory. All this explains our interest
in the $\Box R$ part of the trace anomaly in Weyl gravity theory.

Thus, the derivation of the term $\Box R$ in the one-loop trace
anomaly is the main subject of the present work. As we will see
the calculation of this term requires not only the use of known
algorithms for the minimal fourth-order
operator~\cite{frts82,bavi85,Gusynin90,OP2015}, but also the
generalization of the algorithms for the nonminimal second-order
vector operator~\cite{frts82,bavi85,Gusynin99,OP2015}.

Though the $\Box R$ term was ignored in previous calculations in
conformal gravity~\cite{frts82,antmot,Weyl,OP2015,OP2013}, its
calculation in generic curvature squared non-conformal theory was
successfully accomplished \cite{frts82}. We emphasize again that the
status of this term in non-conformal case is different because the
relevant $R^2$ term in the classical action is not forbidden by the
requirement of local Weyl invariance and is subject to the
renormalization and experimental adjustment in the subtraction point.

The paper is organized as follows. In Sec.~\ref{sec2} we explain
the role of $\Box R$ term in quantum field theory on classical curved
background and in conformal
quantum gravity and emphasize the fundamental difference between
the two cases. In the subsequent sections, we describe the
calculation of the  $\Box R$-type anomaly in the Weyl conformal
gravity. Sec.~\ref{sec3} discusses the background field method in
the conformal quantum gravity, presents the results for Hessian
operators and introduces gauge fixing
condition. On top of this, we repeat the proof of~\cite{a}
of the gauge-fixing independence of the result for the one-loop
divergences in this theory.  Sec.~\ref{ghost-weight} gives
the final forms of Faddeev-Popov ghosts and weight operator
contributing to the one-loop divergences along with the tensor
sector coming from quantum metric.
In  Sec.~\ref{s-totalderdev} we briefly summarize the algorithms
for the heat-kernel-based algorithms for the minimal
fourth-derivative operators acting in space of arbitrary fields~\cite{frts82,bavi85,Gusynin90,OP2015}
and for the nonminimal second-order vector operator~\cite{Gusynin99}.
Most of the calculational efforts of the present work consisted
in the verification of this result by the method of
universal functional traces of~\cite{bavi85}. We report on this extensive
calculation in Sec.~\ref{s-nonmi} and show that the mentioned
algorithm of~\cite{Gusynin99} is confirmed.
Section~\ref{sec5} reports on the final derivation of the anomaly $\Box R$ term by using these algorithms.
In Sec.~\ref{conclusions} we present the conclusions and final discussions of the conformal
symmetry breakdown in Weyl quantum gravity by the \textit{local} term.

Throughout the paper we use Euclidean notations.

\section{Trace anomaly and induced action of gravity}
\label{sec2}

Let us briefly review the conformal anomaly and derivation of the
anomaly-induced action~\cite{rie,FrTs84}.
In the discussion of this subject, we shall
pay special attention to the ambiguities related to the choice of
regularization and to the difference between conformal theories in external gravitational
field and conformal quantum gravity.

The starting classical theory of the fields $\Phi_i$ and the metric,
is conformal, i.e., its action $S_{conf}$ satisfies the conformal
Noether identity, that is a generalization of (\ref{NoetherW}),
\beq
\frac{1}{\sqrt{g}}\,
\bigg\{
2g_{\mu\nu} \frac{\delta }{\delta g_{\mu\nu}}
+ \sum_{i} w_i \Phi_i \frac{\delta }{\delta \Phi_i}\bigg\}\,S
\,=\,0,
\label{Noether_gen}
\eeq
were $w_i$ is the conformal weight of the field $\Phi_i$ and $S$ is
the action of gravity and fields $\Phi_i$. In the purely
gravitational sector, the action has the form (\ref{action conf}),
and the form of conformal actions of scalars, fermions and vectors can be
found, e.g., in~\cite{OUP} or elsewhere. In pure quantum gravity
or on shell of matter fields, when $\de S/\de \Phi_i=0$, the Noether identity (\ref{Noether_gen})
reduces to (\ref{NoetherW}). In this sense, conformal quantum gravity and semiclassical
theories are similar. Another common point is that, at the one-loop level,
the identity (\ref{NoetherW}) gets violated, i.e., acquires a nonvanishing mean value
\beq
\langle T^\mu_{\,\,\mu}\rangle
\,=\, -\, \frac{2}{\sqrt{g}}\,g_{\mu\nu}
\frac{\delta \Ga}{\delta g_{\mu\nu}}\,\neq\,0.
\label{anomaly}
\eeq
In the case of quantum matter fields, the trace anomaly corresponds to the violation of conformal symmetry
in the finite vacuum part of effective action. Depending on the regularization scheme,
the expression for $\langle T^\mu_{\,\,\mu}\rangle$ is proportional to the local one-loop divergences
modulo the $\Box R$ term which is the object of our prime interest. In quantum theory of conformal matter fields,
these divergences and trace anomaly consist of the $C^2$, $E_4$ and $\Box R$~\cite{tmf}. Thus within the existing
classification of invariants~\cite{ddi,DeserSchwimmer}
\beq
\langle T^\mu_{\,\,\mu}\rangle
\,=\,
\om C^2
+ b E_4
+ c \Box R ,
\label{T}
\eeq
where $\om$, $b$ and $c$ depend on the fields content of the model. Equation~(\ref{anomaly}) can be used
to find the part of the full effective action, which is responsible for the anomaly
and also called induced effective action.

The anomaly can be integrated using the relations (\ref{identity-a}) and
$\sqrt{g}\,\big(E_4-\frac23\,{\Box} R\big)
\,=\,
\sqrt{\bar g}\,\big({\bar E_4}-\tfrac23\,\bar\Box {\bar R}
+ 4{\bar \De_4}\si \big)$,
where the metrics $g_{\mu\nu}$ and $\bar g_{\mu\nu}$ are related by $g_{\mu\nu}=\bar g_{\mu\nu} e^{2\sigma}$ and
\beq
\De_4 = \Box^2 + 2\,R^{\mu\nu}\na_\mu\na_\nu - \frac23\,R{\Box}
+ \frac13\,(\na^\mu R)\na_\mu,
\label{119}
\eeq
is the fourth-order Hermitian conformal invariant operator~\cite{FrTs-superconf,Paneitz}
acting on a scalar field of zero conformal weight.
\
The covariant solution of Eq.~(\ref{identity-a}) has the form
\beq
\Gamma_{ind}
&=&
S_c(g_{\mu\nu}) \,+\, \frac{\om}{4}\,\int_x
\int_y\, C^2(x)
\,G(x,y)\,\Big( E_4 - \frac23 \Box R \Big)_y
\nn
&&
\,+\,
\frac{b}{8}\,\int_x \int_y\,
\Big(E_4 - \frac23 \Box R\Big)_x
\,G(x,y)\,\Big( E_4 - \frac23 \Box R \Big)_y
\,-\,
\frac{3c+2b}{36}\,\int_x R^2(x) .
\mbox{\qquad}
\label{abc-terms}
\eeq
Here $S_c(g_{\mu\nu})$ is arbitrary conformally invariant functional of metric, serving as
an integration constant for this solution. This term is important for short distance behavior of stress-tensor
correlators~\cite{Erdmenger-Osborn} or in the model of initial conditions for inflationary cosmology
driven by a conformal field theory~\cite{Hand_BK,debate}, but for the purposes of our paper
it is largely irrelevant (see, e.g., the discussion in~\cite{PoImpo}). The next two terms include
the conformal Green function $G(x,y)$ of the operator $\De_4$ and are free of ambiguities.
On the other hand, the situation is more complex with the local term $\int \sqrt{g}R^2$, which
is directly related to the $\Box R$-type anomaly owing to the relation (\ref{identity-a}).

In semiclassical theories, one can modify $\Box R$-term in
the anomalous trace simply by adding the $\int \sqrt{g}R^2$
term to the classical action. This procedure can be also hidden
in the details of dimensional or Pauli-Villars regularizations~\cite{anomaly-2004}.
However the corresponding ambiguities are in fact equivalent to adding a classical $\int \sqrt{g}R^2$
term. In a semiclassical model, this vacuum term does not
produce changes in the quantum theory because the metric is
not quantized. However, things change if we add
such a term in the theory of conformal quantum gravity, because this
operation violates the classical conformal symmetry, increases
the number of degrees of freedom and changes the quantum theory.

\section{Background field method and gauge fixing}
\label{sec3}

For a one-loop calculation, we shall use the background field method,
as it was done in the previous works in the same model, starting from~\cite{frts82}.
The first step is to separate the metric field into
background $g_{\mu\nu}$ and quantum  $h_{\mu\nu}$ counterparts,
\beq
g_{\mu\nu} \,\longrightarrow\, g'_{\mu\nu} = g_{\mu\nu} + h_{\mu\nu}.
\label{bfm}
\eeq
The one-loop calculations include contribution of the Hessian for the
fluctuation $h_{\mu\nu}$, so we need to expand the action up to
second order in this field. One detail makes our calculation
different from what was done before. We need only the $\Box R$-type term
and, therefore, can restrict our attention to the linear in curvature tensor terms,
since other contributions were calculated and verified in~\cite{frts82,Weyl,OP2015,antmot}.

In what follows, we use standard condensed notations
of DeWitt~\cite{DeWitt65}. The unity operator and the covariant derivatives obeying the Leibnitz rule read as
\bea
\de_{\mu\nu,\a\b}
= \frac12 \big(g_{\mu\a}g_{\nu\b} +g_{\mu\b}g_{\nu\a}\big),
\qquad
\na_\mu A = (\na_\mu A) + A\na_\mu.
\eea

In the framework of usual Faddeev-Popov approach,
we add the background-invariant gauge-fixing term fixing the
diffeomorphism invariance for the quantum fields,
\bea
S_{gf} \,=\, \int d^4 x \sqrt{g} \,\chi_\mu \,Y^{\mu\nu} \chi_\nu,
\label{Sgf}
\eea
with the gauge condition $\chi_\mu$ and the gauge fixing operator $Y^{\mu\nu}$ of the general form
\beq
\chi_\mu = \nabla_\a h^\a_\mu + \tau\, \nabla_\mu h,
\qquad
Y^{\mu\nu} \,=\, \ga_1 g^{\mu\nu} \Box + \ga_2 \nabla^\mu \nabla^\nu,
\label{gf-gen}
\eeq
where $h$ is the trace, $h=g^{\mu\nu}h_{\mu\nu}$, while $\tau$
and $\ga_{1,2}$ are three gauge-fixing parameters. For arbitrary
choice of these parameters, the calculations become complicated.
The total Hessian contributed by the sum of the action (\ref{action conf})
and the gauge-fixing term (\ref{Sgf}) is a nonminimal fourth-order operator.
On the other hand, there is a unique special choice of the gauge-fixing parameters~\cite{Weyl,OP2013,OP2015}.
\beq
\ga_1  = \frac12,
\qquad
\ga_2 = -\frac16,
\qquad
\tau = - \frac14,
\label{gauge-min}
\eeq
that reduces this Hessian to the minimal form. Then the quadratic form of
the action reads as
\bea
S_{conf}^{(2)} = \frac14 \int d^4 x \sqrt{g}\, h^{\mu\nu}
H_{\mu\nu,\a\b} h^{\a\b},
\eea
where
\bea
\hat{H} \, = \, H_{\mu\nu,\a\b}
\, = \,  \hat{1} \Box^2 + \hat{V}^{\rho\si} \na_\rho\na_\si
+ \hat{N}^\rho \na_\rho +\hat{U}.
\label{Hmingen}
\eea

At this point, we encounter a special feature of conformal gravity,
which has not only diffeomorphism invariance, but also conformal symmetry.
For the gauge fixing of conformal symmetry we follow~\cite{frts82,OP2015}
and choose the degenerate (delta function type) gauge $\phi\equiv g^{\mu\nu}h_{\mu\nu}=0$.
This means that the quantum metric fluctuation in the path integral $h_{\mu\nu}$ becomes traceless and the unity
matrix in the metric sector is a projector operator to the subspace of traceless tensors
(the hat indicating the action of the tensor matrix objects in the space of tensors)
\bea
\hat{1} = \d_{\mu\nu,\a\b}-\frac14 g_{\mu\nu} g_{\a\b}.
\label{traceless}
\eea

After some algebra, we get 
the following elements of the operator~(\ref{Hmingen}):
\beq
\hat{V}^{\rho\sigma}
&=&
- \frac23 R \d_{\mu\nu,\a\b} g^{\rho\si}
+\frac43 R g_{\nu\b} \delta_{\mu\a,}^{\ \ \ \rho\si}
+\frac43 R_{\a\b}  \delta_{\mu\nu,}^{\ \ \ \rho\si}
+ \frac43 R_{\mu\nu} \delta_{\a\b,}^{\ \ \ \rho\si}
\nn
&&
+ \, 2 R_{\mu\a} \delta_{\nu\b,}^{\ \ \ \rho\si}
- 4 R_{\mu}^\rho g_{\nu\b} \d_\a^\si
- 4 R_{\a}^\rho g_{\nu\b} \d_\mu^\si
+ 4 R_{\mu\a\nu\b} g^{\rho\si}
+ 2 \d_{\mu\nu,\a\b} R^{\rho\si},
\label{Vrhosi}
\eeq
\beq
\hat{N}^\lambda
&=&
\frac13\d_{\mu\nu,\a\b} \pare{\nabla^\la R}
-\frac43 \pare{\nabla_\mu R_{\a\b}} \d_\nu^\la
-\frac23 \pare{\nabla_\a R} g_{\nu\b} \d_\mu^\la
- 2 \pare{\nabla_\mu R_{\nu\b}} \d_\a^\la
\nn
&&
+ 4 \pare{\nabla_\a R_{\mu\nu}} \d_\b^\la
+ 4 \pare{\nabla_\a R_{\mu\b}} \d_\nu^\la
- 4 \pare{\nabla_\a R_{\mu}^\la} g_{\nu\b}
+ 4 \pare{\nabla^\la R_{\mu\a\nu\b}},
\label{Nrho}
\eeq
and
\beq
\hat{U}
&=&
- \frac13 \d_{\mu\nu,\a\b} \pare{\Box R}
-\frac43 \pare{\nabla_\mu \nabla_\a R} g_{\nu\b}
+\frac43 \pare{\nabla_\mu \nabla_\nu R_{\a\b}}
\nn
&&
+ \,2 \pare{\Box R_{\mu\a}} g_{\nu\b}
+ 2 \pare{\Box R_{\mu\a\nu\b}}.
\label{U}
\eeq

Before starting the calculations, let us comment on the important question about the gauge-fixing dependence.
Repeating the arguments of~\cite{a,OUP} (and those of~\cite{frts82} for a general higher-derivative quantum
gravity) this issue may be stated as follows.
As is well known, the gauge and parametrization dependence of the one-loop
effective action is proportional to classical equations of
motion~\cite{aref,Weinberg,Costa-Tonin,volatyu}. In view of locality of UV divergences~\cite{Collins,Lavrov-renQG}
in the theory~(\ref{action conf}), the difference between its divergences $\Delta\Ga^{(1)}_{div}$ calculated
in different {\em background-covariant} gauges~\cite{BRST}, which necessarily has dimensionality 4
and is proportional to equations of motion, can only be of the following form~\cite{a}
\beq
\Delta\Ga^{(1)}_{div}\propto
\int d^4x\,\,
g_{\mu\nu}\, \, \frac{\de S_W}{\de g_{\mu\nu}}\label{conf-amb}
\eeq
(other powers of $\de S_W/\de g_{\mu\nu}$ are obviously excluded for dimensional reasons).
Therefore, in view of the conformal symmetry of the classical action (\ref{NoetherW}), it is vanishing.
Thus, in the conformal theory  (\ref{action conf}) in this class of gauges the one-loop divergences
are parametrization and gauge-fixing independent. Hence we can safely use the simplest choices of
variables and gauge-fixing conditions.

\section{Gauge ghosts and gauge-fixing operator}
\label{ghost-weight}

The general expression for the one-loop effective action in Euclidean notations
reads~\cite{frts82} (see also~\cite{avbar86,Ohta2020} and the recent review in~\cite{Hand-NO})
\beq
{\Ga}
\,=\, \frac{1}{2}\,\Tr\log \hat{H}
\,-\, \frac{1}{2}\,\Tr\log \hat{Y}
\,-\, \Tr\log \hat{M},
\label{one-loop-divs}
\eeq
where $\Tr$ denotes the functional trace, $\hat{H}$ is defined in (\ref{Hmingen}), the gauge-fixing operator
\beq
\hat{Y} = Y^{\mu\nu}
\,=\, \frac12\Big( g^{\mu\nu} \Box - \frac13\, \na^\mu \na^\nu\Big),
\label{Y}
\eeq
is defined in (\ref{gf-gen}) and (\ref{gauge-min}), and $\hat{M} = M^\be_{\,\,\al}$ is the Hessian
of the action of ghost fields responsible for diffeomorphism gauge transformations,
\beq
M_{\be\al}
\,=\, \frac{\de \chi_\be}{\de h_{\mu\nu}}\,{\cal R}_{\mu\nu,\al},
\label{FPghosts}
\eeq
with the generator of diffeomorphism transformations
${\cal R}_{\mu\nu,\al} = - g_{\mu\al}\na_\nu  - g_{\nu\al}\na_\mu$.
This operator is built with the diffeomorphism gauge conditions in
(\ref{gf-gen})-(\ref{gauge-min}). Regarding the conformal gauge $h=0$ we note that its ghost field
does not contribute to the effective action~(\ref{one-loop-divs}) owing to the non-dynamical (non-derivative)
nature of its conformal ghost sector.\footnote{This is a nontrivial and very helpful corollary
of the choice of the diffeomorphism gauge (\ref{gf-gen})-(\ref{gauge-min}). In the full set of
diffeomorphism and conformal gauge conditions $(\chi_\mu,\phi)$, $\phi\equiv g^{\mu\nu}h_{\mu\nu}$,
the full Faddeev-Popov operator has a block matrix form with the off-diagonal element
$(\delta\chi_\alpha/\delta h_{\mu\nu}){\cal R}_{\mu\nu}$, where ${\cal R}_{\mu\nu}=g_{\mu\nu}$ is
the linearized generator of the conformal transformation of $h_{\mu\nu}$. But in the gauge (\ref{gf-gen})
with the parameter $\tau=-1/4$ of (\ref{gauge-min}) this block turns out to be vanishing, so that
the total ghost determinant factorizes into the product of ${\rm Det}\,M_{\alpha\beta}$ and the determinant
of its conformal-conformal block $(\delta\phi/\delta h_{\mu\nu}){\cal R}_{\mu\nu}=4$.
The latter one is ultralocal and does not contribute to the effective action.} Thus we get
\beq
\hat{M} \,=\, M^\be_{\,\,\al}
\,=\,
-\Big(
\de^\be_{\,\,\al}\Box + \frac12\,\na^ \be \na_\al + R^\be_{\,\,\al}
\Big).
\label{FPop}
\eeq
In both cases of operators (\ref{Y}) and (\ref{FPop}) we meet the problem of calculating $\Tr\log \hat{F}$
for the nonminimal vector operator
\beq
\hat{F}\,\equiv \,
F^\mu_\nu=\Box\delta^\mu_\nu-\la\nabla^\mu\nabla_\nu+P^\mu_\nu
\label{vec.nomin}
\eeq
with a generic value of the parameter $\lambda$ and some potential term $P^\mu_\nu$.

\section{The algorithms for anomalies and divergences}
\label{s-totalderdev}

One-loop UV divergences and anomalies in curved spacetime can be covariantly calculated by using the heat kernel
of the corresponding wave operators of the theory. For the effective action built in terms of the functional
determinants of the second-order operators $\hat F$ and the fourth-order operators $\hat H$, the heat kernels
of these operators can be represented in the form of asymptotic expansions at small values of the proper time $s\to 0$:
\beq
e^{s\hat F}\delta(x,y)\,\Big|_{\,y=x}=\frac{\sqrt{g(x)}}{(4\pi s)^{d/2}}\sum_{m=0}^\infty s^m\hat a^F_m(x),
\eeq
and
\beq
e^{s\hat H}\delta(x,y)\,\Big|_{\,y=x}=\frac{\sqrt{g(x)}}{(4\pi s^{1/2})^{d/2}}\sum_{m=0}^\infty s^{m/2}\hat
a^H_m(x)
\eeq
where $d$ is the spacetime dimension, and $\hat a^{F,H}_{m}(x)$ are respectively the local
Schwinger-DeWitt~\cite{DeWitt65,bavi85} coefficients of the operators $\hat F$ and $\hat H$ related to
the Gilkey-Seeley coefficients~\cite{Gilkey} $\hat E^{F,H}_{2m}(x)$ by
\beq
\hat a_m^{F,H}(x)={(4\pi)^{d/2}}\,\hat E^{F,H}_{2m}(x), \label{SDW-GS}
\eeq

These heat kernels generate by integration over the proper time parameter the contributions to the Euclidean
effective action $\Gamma=\tfrac12{\rm Tr}\ln F=\tfrac12\int_0^\infty ds\,e^{-sF}/s$ coming from the operator
$\hat F$ and correspondingly from $\hat H$. Their divergent parts read respectively as
\beq
\frac{1}{2}\,\Tr\log \hat F\,\Big|_{\,div}=-\frac1\epsilon\int d^4x\,\sqrt{g}\,{\rm tr}\,\hat a_2^F, \label{Fdiv}\\
\frac{1}{2}\,\Tr\log \hat H\,\Big|_{\,div}=-\frac2\epsilon\int d^4x\,\sqrt{g}\,{\rm tr}\,\hat a_2^H, \label{Hdiv}
\eeq
where $\rm tr$ denotes the matrix trace with respect to the indices of the second Schwinger-DeWitt coefficient
$\hat a_2$ and $\ep = (4\pi)^2(4-n)\to 0$ is the parameter of dimensional
regularization with $n$ denoting the regularized spacetime dimension. To avoid the ambiguities related
to this regularization~\cite{Duff94,anomaly-2004}, one can use the
covariant cutoff in the proper time regularization with the dimensionless parameter
$L\to\infty$ representing the ratio of the dimensional UV cutoff and the renormalization scale (see~\cite{bavi85})
\beq
\frac{1}{\ep}\,=\,\frac{\log L^2}{32\pi^2}.
\label{logL}
\eeq
Note the difference in coefficients in (\ref{Fdiv}) and (\ref{Hdiv}) associated with the fact that
the operator $\hat H$ is quartic in derivatives (that is, it has the dimension four in units of mass and
the corresponding conformal weight $-4$), whereas weight $-2$ operator $\hat F$ is quadratic in derivatives.
This weight plays important role in the generation of trace anomaly, which can be most easily demonstrated
within zeta-functional regularization as follows.

Consider a symmetric Weyl-covariant operator $\hat H$ of the conformal weight $-2w$ (correspondingly
the weight of the field acted upon by $\hat H$ being $w$), which transforms under the infinitesimal conformal
rescaling $\delta_\sigma g_{\mu\nu}=+2\sigma g_{\mu\nu}$ as $\delta_\sigma\hat H=-w(\sigma\hat H+\hat H\sigma)$.
The conformal transform of its ``effective action'' $\Gamma=\tfrac12\log\hat H$, expresses in zeta function
regularization in terms of the particular value of the zeta function,
$\zeta(z|x)={\rm tr} (-\hat H)^{-z}\delta(x,y)\,\big|_{\,y=x}$, at $z=0$ --- the value which is given
in 4-dimensional massless theories in terms of the local Schwinger-DeWitt (or Gilkey-Seeley) coefficient
(\ref{SDW-GS}), $\zeta(0|x)={\rm tr}\,\hat E^H_4(x)$. Namely,
\beq
\langle \, T^\mu_\mu\,\rangle\equiv-\delta_\sigma\Gamma =w\,\zeta(0|x)=w\,{\rm tr}\,\hat E^H_4(x).
\eeq
Thus, when the both operators $\hat F$ and $\hat H$ are individually Weyl invariant with respect to
local conformal transformations, their functional determinants supply the total trace anomaly with
the following contributions expressible in terms of the second Schwinger-DeWitt coefficient
\beq
\label{FHanomaly}
\langle\, T^{\mu\,(F)}_\mu\rangle=\frac{{\rm tr}\,\hat a_2^F(x)}{16\pi^2},\quad
\langle\, T^{\mu\,(H)}_\mu\rangle=2\frac{{\rm tr}\,\hat a_2^H(x)}{16\pi^2},
\eeq
in accordance with their conformal weights $-2w=-2$ and $-2w=-4$ respectively.

In Weyl gravity theory the fourth-order operator $\hat H$ given by (\ref{Hmingen}) and nonminimal
second-order operators $\hat F=(M_{\mu\nu},Y^{\mu\nu})$ given by Eqs.(\ref{Y})-(\ref{FPghosts}) are
not individually Weyl covariant, because the chosen diffeomorphism and conformal gauges are not background
covariant with respect to the local conformal subgroup of the full set of gauge transformations.
However, these operators are intertwined by Ward identities which provide the expression (\ref{conf-amb})
for gauge conditions variation of the one-loop divergences, which in its turn occurs to be vanishing also
off shell in view of the discussion above. Thus the deviation of the actual conformal transformations from
the expressions (\ref{FHanomaly}) above in fact cancels out in the total sum of contributions into effective
action (\ref{one-loop-divs}), which generate the overall trace anomaly
\beq
\label{FHanomaly2}
\langle\, T^{\mu}_\mu(x)\rangle=\frac1{16\pi^2}\Big(2\,{\rm tr}\,\hat a_2^H(x)-
{\rm tr}\,\hat a_2^Y(x)-2\,{\rm tr}\,\hat a_2^M(x)\Big).
\eeq
This expression thus turns out to be the integrand of the overall set of the one-loop divergences of the model,
\beq
\label{FHdiv}
\Ga\,\big|_{\,div}=-\frac1\epsilon \int d^4 x \sqrt{g}\,
\big(2\,{\rm tr}\,\hat a_2^H-
{\rm tr}\,\hat a_2^Y-2\,{\rm tr}\,\hat a_2^M\big).
\eeq
Note, however, that unless we include the surface terms at the boundary of spacetime, which are disregarded
in this four-dimensional integral, the $\Box R$ contributions -- the object of our prime interest -- are
completely washed out from the right-hand side of (\ref{FHdiv}) because they do not contribute to
the interior of spacetime domain. But they enter the right-hand side of (\ref{FHanomaly2}).
They are ambiguous, as it was already remarked, in the dimensional regualarization, but in the zeta-functional
regularization appear as uniquely defined ingredients of the relevant Schwinger-DeWitt (or Gilkey-Seeley)
coefficients. Our goal is to calculate them here.

Let us start the calculation of various terms of (\ref{FHanomaly2}) with the contribution of the minimal
fourth-derivative operator (\ref{Hmingen}). We shall ignore in what follows all the terms which are
not of the desired $\Box R$ type, as they are not of our interest. The reader can easily find
the corresponding expressions in~\cite{frts82,bavi85} or in~\cite{Gusynin90,Gusynin99,Barv-Wach}.
The expression for the $\Box R$-type terms for the fourth order minimal operator was derived
in~\cite{Gusynin90} using the heat-kernel method. The formula obtained in~\cite{Gusynin90} is equivalent,
for the operator (\ref{Hmingen}), to
\beq
{\rm tr}\,\hat a_2^H(x)
=\frac12\tr \Big\{
\frac{\hat{1}}{15}\pare{\Box R}
+\frac19\,\Box\hat{V}
-\frac{5}{18}\,\nabla_\rho\nabla_\si \hat{V}^{\rho\si}
+\frac12\, \nabla_\la\hat{N}^\la - \hat{U}\Big\},
\label{Gusyn90}
\eeq

It is worth mentioning that this expression passes the test of representing the fourth-order derivative
operator as a product of two minimal second order ones~\cite{frts82} and calculating the divergent part
of its functional determinant. The details of this verification sound as follows. For the basic second-order
operator, the divergent part is~\cite{DeWitt65}
\beq
- \frac{1}{2}\,\Tr\log \big(\hat{\Box} + \hat{\Pi}\big)\,\Big|_{\,div}
&=&
- \frac{1}{\ep}
\int d^4 x \sqrt{g}\,\tr \Big\{
\frac{\hat{1}}{30}\,\Box R
\,+\,\frac{\hat{1}}{6}\,\Box \hat{\Pi} \Big\},
\label{basic}
\eeq
Taking the particular form of the fourth-derivative operator
\beq
\hat{O} &=& \big(\hat{\Box} + \hat{\Pi}\big)^2
\,=\, \hat{\Box}^2 + 2\hat{\Pi}\Box
+ 2\big(\na^\la \hat{\Pi}\big)\na_\la
 + \big(\Box \hat{\Pi}\big)\,,
\nn
&&
\Tr\log \hat{O} \,=\, 2 \Tr\log \big(\hat{\Box} + \hat{\Pi}\big),
\label{OPi2}
\eeq
we identify this with the special version of (\ref{Hmingen}), when
\beq
\hat{V}^{\rho\si} \,=\, 2\hat{\Pi}g^{\rho\si},
\quad
\hat{N}^\rho \,=\, 2\big( \na^\rho \hat{\Pi}\big),
\quad
\hat{U} \,=\, \big( \Box \hat{\Pi}\big).
\label{VNU}
\eeq

At this point, we can establish the form of the possible divergences
of the $\Box R$-type. These divergences should be the total derivative
expressions constructed from $\hat{V}^{\rho\si}$, $\hat{N}^\rho$
and $\hat{U}$. taking into account the dimension of these building
blocks, we arrive at the expression for divergences which has four
unknown coefficients $a_{1,..,4}$, that are still to be defined,
\beq
\frac{1}{2}\,\Tr\log \hat{H}\Big|_{div}
&=&
-\frac{1}{\ep}
\int d^4 x \sqrt{g}\,\tr \Big\{
\frac{\hat{1}}{90}\,\Box R
+ a_1\na_\rho \na_\si \hat{V}^{\rho\si}
\nn
&&
+\,\, a_2\Box \hat{V}
+ a_3 \na_\rho \hat{N}^\rho
+ a_4 \hat{U}\Big\},
\label{a1234}
\eeq
From the non-total derivative terms we know that $a_4=-1$ (e.g.,~\cite{frts82,bavi85}).
Other coefficients can be obtained using different doubling tricks~\cite{frts82}.
In this case, we can identify
$\na_\rho \na_\si \hat{V}^{\rho\si} = 2\Box \hat{\Pi}$
and $\Box \hat{V} = 8\Box \hat{\Pi}$. Using these relations and
(\ref{OPi2}), we arrive at the equation
\beq
2a_1 + 8a_2 + 2a_3 + a_4 \,=\,\frac13.
\label{conda1234}
\eeq
Since Eq.~(\ref{Gusyn90}) fits this condition, we conclude that it
has passed this partial check.

Let us now consider the contribution of the nonminimal vector
operator (\ref{vec.nomin}). The algorithms for the divergences
in this case is known from~\cite{frts82} and~\cite{bavi85}. However,
in both cases the formula for divergences did not include the total
derivative terms, such as $\Box R$,  $\Box P$  and
$\na_\mu \na_\mu P^{\mu\nu}$, where $P = P^{\mu\nu}g_{\mu\nu}$.
The algorithm for these terms was obtained in~\cite{Gusynin99}.
The final result for the vector operator (\ref{vec.nomin}),
$\hat a_2^F\equiv (a_2^F)^\mu_\nu$, ${\rm tr}\,\hat a_2^F=(a_2^F)^\mu_\mu$, reads
\beq
{\rm tr}\,\hat a_2^F
&=&
c_1 \Box P
+ c_2 \nabla^\a\nabla^\b P_{\a\b}+ c_{11}\Box R,
\label{nonmin-GK}
\eeq
with the coefficients
\beq
c_1 &=&
- \frac{8\la^2-21\la+6}{36\la(1-\la)}
 +  \frac{2\la-1}{6\la^2}\log(1-\la),
\nn
c_2
&=&
- \frac{13\la^2+6\la-24}{36\la(1-\la)}
+ \frac{\la+4}{6\la^2}\log(1-\la),
\nn
c_{11}
&=&
- \frac{133\la^2-168\la-60}{360\la(1-\la)}
- \frac{\la^2-5\la-2}{12\la^2}\log(1-\la),
\label{GuKo-E4}
\eeq
where we keep enumeration of the coefficients adopted in~\cite{Gusynin99}.
To have an additional verification, in the next section we present an alternative derivation
of the contribution of minimal-vector operator, by using a different approach.

\section{New derivation for nonminimal vector operator}
\label{s-nonmi}

Consider the vector field operator (\ref{vec.nomin})
\begin{eqnarray}
F\equiv F^\mu_\nu
=\Box\delta^\mu_\nu-\la\nabla^\mu\nabla_\nu+P^\mu_\nu .
\end{eqnarray}
Coincidence limit of the heat kernel has
Schwinger-DeWitt expansion given by~\cite{DeWitt65,bavi85}
\beq
e^{sF}\delta^\mu_\nu(x,y)\,\Big|_{\,y=x}
= \frac1{(4\pi s)^{d/2}}\,g^{1/2}
\Big[
a^\mu_{0\,\nu}(x,x) + a^\mu_{1\,\nu}(x,x) s +a^\mu_{2\,\nu}(x,x)s^2
+ \,... \Big],
\label{Sch-Dew}
\eeq
where the $a^\mu_{n,\nu}(x,x)$ are the coincidence limits of the two-point Schwinger-DeWitt
coefficients $a^\mu_{n,\nu}(x,y)$ labeled above by one argument $a^\mu_{n,\nu}(x)\equiv a^\mu_{n,\nu}(x,x)$.

We intend to find total derivative terms in the second
Schwinger-DeWitt coefficient $a^\mu_{2\,\nu}(x,x)$ which
determine the corresponding one-loop divergences in the
effective action. For the matrix trace of $a^\mu_{2\,\nu}(x,x)$
there are three such structures with some numerical coefficients
$a,b$ and $c$,
\beq
a^\mu_{2\,\mu}(x,x)
= a\Box R+b\Box P+c\nabla_\mu\nabla^\nu P^\mu_\nu,
\quad
P\equiv P^\mu_\mu.
\end{eqnarray}

These terms cannot be extracted from the integral quantity like
${\rm Tr}\,e^{sF}$ or ${\rm Tr}\,\log F$ because under integration
over spacetime they get washed out and materialize as surface
terms at the boundary which we do not control. Therefore, let
us extract them from the local quantity. The simplest object is the
coincidence limit of the Green's function
\beq
-\frac1F\delta^\mu_\nu(x,y)\,\Big|_{\,y=x}=\int\limits_{0}^\infty ds\,e^{sF}\delta^\mu_\nu(x,y)\,\Big|_{\,y=x}.
\eeq
Unfortunately, however, $a^\mu_{2\,\nu}(x,x)$ is contained in the
UV finite part of this quantity which for massless operator is
badly IR divergent within the local Schwinger-DeWitt expansion~(\ref{Sch-Dew}).
Therefore we have to consider the massive Green's function
\beq
&&
\frac1{m^2-F}\delta^\mu_\nu(x,y)\,\Big|_{\,y=x}
= \int_{0}^\infty ds\,e^{sF-sm^2}\delta^\mu_\nu(x,y)\,\Big|_{\,y=x}
\nn
&&
\quad
= \int_{0}^\infty\frac{ds\,e^{-sm^2}}{(4\pi s)^{d/2}}\,g^{1/2}
\Big[a^\mu_{0\,\nu}(x,x) + a^\mu_{1\,\nu}(x,x)\,s
+ a^\mu_{2\,\nu}(x,x)\,s^2+...\Big]
\nn
&&
\quad
= \frac1{(4\pi)^{d/2}}\,g^{1/2}
 \bigg[
 \frac{\varGamma(1-\frac{d}2)}{(m^2)^{1-\frac{d}2}}
 \,a^\mu_{0\,\nu}(x,x)
 + \frac{\varGamma(2-\frac{d}2)}{(m^2)^{2-\frac{d}2}}
 \,a^\mu_{1\,\nu}(x,x)\,
 +\frac1{m^2}\,a^\mu_{2\,\nu}(x,x)+...\bigg],\qquad
\label{heatk}
\end{eqnarray}
for $d\to 4$. Then $a^\mu_{2\,\nu}(x,x)/16\pi^2$ is the coefficient
of $1/m^2$ in the inverse mass expansion of a massive Green's
function.

Let us calculate this Green's function by the method of universal functional traces of~\cite{bavi85}.
In the lowest order in curvatures, we have
\beq
\frac1{F-m^2}
=K^\mu_\a\frac{\delta^\a_\nu}{(\Box-m^2)\,(\Box-\frac{m^2}{1-\la})}+...\,,
\eeq
where
\begin{eqnarray}
K\equiv K^\mu_\nu=\Big(\Box-\frac{m^2}{1-\la}\Big)\,\delta^\mu_\nu
+\frac\la{1-\la}\,\nabla^\mu\nabla_\nu.
\label{7}
\end{eqnarray}
More precisely, we can derive the exact equality:
\begin{eqnarray}
(F^\mu_\a-m^2\delta^\mu_\a)\,K^\a_\nu
=(\Box-m^2)\,\Big(\Box-\frac{m^2}{1-\la}\Big)\,\delta^\mu_\nu+M^\mu_\nu,
\label{8}
\end{eqnarray}
where the perturbation operator $M^\mu_\nu$ equals
\begin{eqnarray}
&&M^\mu_\nu=\frac\la{1-\la}\,(R^\mu_\a
+P^\mu_\a)\nabla^\a\nabla_\nu
+P^\mu_\nu\,\Big(\Box-\frac{m^2}{1-\la}\Big)-\la R_{\nu\a}
\nabla^\mu\nabla^\a
\nn
&&
\qquad\qquad-\la(\nabla^\mu R_{\nu\a})\nabla^\a
-\frac\la2(\nabla_\nu R)\nabla^\mu
-\frac\la2(\nabla^\mu\nabla_\nu R).
\label{9}
\end{eqnarray}

From (\ref{8}), we find
\begin{eqnarray}
\frac{1}{F-m^2} = K \frac{1}{F_1 F_2+M}.
\label{10}
\end{eqnarray}
where we have introduced the abbreviations
\begin{eqnarray}
F_1=\Box-m^2,\quad F_2=\Box-\frac{m^2}{1-\la}.
\label{11}
\end{eqnarray}
We expand \eqref{10} in the inverse powers of $F_1 F_2$:
\begin{eqnarray}
\frac1{F-m^2}=K\,\sum\limits_{p=0}^\infty(-1)^p\,M_p
\frac1{\big( F_1 F_2\big)^{p+1}}.
\label{12}
\end{eqnarray}
We then find
\begin{eqnarray}
M_0=1,\qquad M_1=M \qquad
M_{p+1}=M\,M_p+\Big[\,F_1 F_2,M_p\,\Big].
\end{eqnarray}
For our purposes of finding linear $\nabla\nabla R$ and $\nabla\nabla P$
terms, we need the above operators up to $p=3$ with
\begin{eqnarray}
M_2
&=&
M M_1+[F_1 F_2,M_1] =  M^2 + \Big[ F_1 F_2 ,M \Big]
\nn
&=&
[\Box,M]\,(\, F_1 +F_2) +\big[\,\Box,[\Box,M]\,\big]+...,
\end{eqnarray}
where only $O(R)$ terms are kept.
We also have
\begin{eqnarray}
M_3 = M M_2+[F_1 F_2,M_2] =  \Big[ F_1 F_2, M_2 \Big] +\cdots
= \big[\,\Box,[\Box,M]\,\big]\,( F_1+F_2)^2 + \cdots,
\label{18}
\end{eqnarray}
where we keep only those terms linear in the curvature with two derivatives.
Then the single and double commutators of $\Box$ with $M$ are also needed:
\begin{eqnarray}
&&
[\,\Box,M^\mu_\nu\,]
= \frac{2\la}{1-\la} \Big[\nabla_\b(R^\mu_\a + P^\mu_\a)\Big]
\nabla^\b\nabla^\a\nabla_\nu
+ 2\big(\nabla^\a P^\mu_\nu\big)\nabla_\a F_2
\nn
&&
\qquad\qquad
- \, 2\la\big(\nabla_\b R_{\nu\a}\big) \nabla^\b\nabla^\mu\nabla^\a
+ \frac{\la}{1-\la} \Big[\Box(R^\mu_\a+P^\mu_\a)\Big]\na^\a\na_\nu
+ \big(\Box P^\mu_\nu\big)\, F_2
\nn
&&
\qquad\qquad
- \la\big(\Box R_{\nu\a}\big) \nabla^\mu\nabla^\a
- 2\la\big(\nabla_\b \nabla^\mu R_{\nu\a}\big) \nabla^\b\nabla^\a
 - \la\big(\nabla_\a\nabla_\nu R\big)\nabla^\a\nabla^\mu
+...\, ,\quad
\eeq
\beq
&&\Big[\,\Box,[\,\Box,M^\mu_\nu\,]\,\Big]
= \frac{4\la}{1-\la} \big[\nabla_\gamma\nabla_\b(R^\mu_\a
+ P^\mu_\a)\big]\nabla^\gamma\nabla^\b\nabla^\a\nabla_\nu
 + 4\big(\nabla^\b\nabla^\a P^\mu_\nu\big)\nabla_\b\nabla_\a F_2
 \nn
&&\qquad\qquad
- 4\la\big(\nabla_\gamma\nabla_\b R_{\nu\a}\big)
 \nabla^\gamma\nabla^\b\nabla^\mu\nabla^\a+... ,
\end{eqnarray}
up to terms $\mathcal{O}(R^2)$, $\mathcal{O}(RP)$ and higher derivatives of the curvatures.

\subsection{$M_0$-term}

Let us explicitly calculate the contribution of the first term in (\ref{12}):
\beq
&& \hs{-10} K^\mu_\a\frac{\delta^\a_\nu}{ F_1 F_2}\,
 \delta(x,y)\,\Big|_{y=x} = \frac{\delta^\mu_\nu}{F_1}\delta(x,y)\, \Big|_{y=x}
+\frac\la{1-\la}\nabla^\mu\nabla_\a \frac{\delta^\a_\nu}{F_1 F_2}
 \delta(x,y)\,\Big|_{y=x}
\label{21}
\eea
The $1/m^2$ term from the first term can be obtained from \eqref{heatk} and is given by
\bea
-g^{1/2}\frac{a_{2\;\nu}^{\Box\,\mu}(x,x)}{16\pi^2\,m^2}
\eea
That from the second term in \eqref{21} is calculated as
\begin{eqnarray}
&&\frac\la{1-\la}\nabla^\mu\nabla_\a
 \frac{\delta^\a_\nu}{(\Box-m^2)\,(\Box-\frac{m^2}{1-\la})} \delta(x,y)\,\Big|_{y=x}\nn
&& = \frac{\la\,g^{1/2}}{1-\la}\nabla^\mu\nabla_\a\int_0^\infty ds\,dt\,\exp\Big[\,(s+t)\Box-m^2
 \Big(s+\frac{t}{1-\la}\Big)\Big] \delta(x,y)\,\Big|_{y=x}\nn
&&=\frac{\la\, g^{1/2}}{1-\la}\int_0^\infty \frac{d\gamma\,\gamma\,
 e^{-m^2\frac{1-\a\la}{1-\la}}}{(4\pi\gamma)^{d/2}}\int_0^1 d\a\,\Big(\nabla^\mu-\frac{\sigma^\mu}{2\gamma}\Big)\,
 \Big(\nabla_\a-\frac{\sigma_\a}{2\gamma}\Big) \nn
&& \hs{20}  \times \Delta^{1/2}\Big(\,a^{\Box\,\a}_{0\,\nu}
 +a^{\Box\,\a}_{1\,\nu}\,\gamma +a^{\Box\,\a}_{2\,\nu}\,\gamma^2+...\Big)\,\Big|_{y=x},
\label{21a}
\end{eqnarray}
where we have used proper time representations for both massive Green's functions ($s$- and $t$-integrals)
and made a change of integration variables $s,t\to\gamma,\a$, $s=\a\gamma$, $t=(1-\a)\gamma$,
with $0\leq s+t=\gamma<\infty$. We have also used the Schwinger-DeWitt expansion~\eqref{Sch-Dew2}
and pulled $\exp(-\sigma(x,y)/2\gamma)$ to the left through two covariant derivatives.
Here $\sigma_\a=\nabla_\a\sigma(x,y)$ and similarly $\sigma^\mu$.

Only two terms containing $\nabla\nabla R$ survive here after differentiation and
taking the coincidence limits. These are where $\nabla^\mu$ acts on $\sigma_\a$ times $a_2^\Box$
(they give $(\nabla^\mu\sigma_\a)a^{\Box\;\a}_{2\;\nu}|_{\,y=x} =a^{\Box\;\mu}_{2\;\nu}(x,x)$)
and where both derivatives act on $a_1^\Box$. They have a needed dimensionality, and contain the needed
$\nabla\nabla R$ terms. Taking in these two terms the integral over $\gamma$ (which can be done directly
in $d=4$, because these terms are UV finite)  we get
\begin{eqnarray}
&&\frac\la{1-\la}\nabla^\mu\nabla_\a \frac{\delta^\a_\nu}{(\Box-m^2)\,(\Box-\frac{m^2}{1-\la})}
 \delta(x,y)\,\Big|_{y=x}\nn
&&\quad= \la \frac{g^{1/2}}{16\pi^2m^2} \int_0^1\frac{d\a}{1-\a\la}\, \Big[\,\nabla^\mu\nabla_\a
 \,a_{1\;\nu}^{\Box\,\a}(x,y) -\frac12\,a_{2\;\nu}^{\Box\,\mu}(x,y)\,\Big]\, \Big|_{\,y=x}+... \nn
&& =- g^{1/2}\frac{\log(1-\la)}{16\pi^2\,m^2}\,
 \Big[\,\nabla^\mu\nabla_\a
 \,a_{1\;\nu}^{\Box\,\a}(x,y)
 -\frac12\,a_{2\;\nu}^{\Box\,\mu}(x,y)\,\Big]\, \Big|_{\,y=x}+...\,.~~~
\end{eqnarray}

Using the coincidence limits (4.32) and (4.33) of~\cite{bavi85} for
the vector field Schwinger-DeWitt coefficients  $a_{n\;\nu}^{\Box\,\a}(x,y)$,
\ $n=0, 1, ...$, of the operator $\Box\delta^\mu_\nu$:
\begin{eqnarray}
\nabla^\mu\nabla_\a \,a_{1\;\nu}^{\Box\,\a}(x,y)\,\Big|_{\,y=x}
&=&
-\frac1{15}\Box R^\mu_\nu+\frac2{15}\,\nabla^\mu\nabla_\nu R+...\,,
\\
a_{2\;\nu}^{\Box\,\mu}(x,x)
&=&
\frac1{30}\,\Box R\,\delta^\mu_\nu+...\,,
\end{eqnarray}
one finally has
\begin{eqnarray}
&&
K^\mu_\a\frac{\delta^\a_\nu}{ F_1 F_2}\,
\delta(x,y)\,\Big|_{y=x}
\,=\,
\frac{g^{1/2}}{16\pi^2\,m^2}\, \Big\{
\Big[-\frac1{30} +\frac1{60}\log(1-\la)\Big]
\,\Box R\,\delta^\mu_\nu
\nn
&&
\qquad
\qquad
\qquad\qquad
- \frac2{15}\, \log(1-\la)\,\nabla^\mu\nabla_\nu R
+\frac{1}{15} \Box R^\mu_\nu\log(1-\la) + ...\,\Big\}.
\label{25}
\end{eqnarray}

\subsection{$M_1$-term}

The rest of the terms can be calculated by using the formulas given
in Appendix~\ref{formula}.
Using (\ref{7}) and (\ref{9}) and retaining only the terms with
$\nabla^2R$ and $\nabla^2P$, we have, for the second term in
\eqref{12},
\beq
&&
- (KM)^\mu_\nu\frac1{F_1^2F_2^2}
\,=\,
g^{1/2} \Big\{- \Big[-\frac\la2\nabla^\mu\nabla_\nu R+\Box P^\mu_\nu+\frac\la{1-\la}
\nabla^\mu\nabla_\a P^\a_\nu\Big]\,\frac1{F_1^2F_2}
\nn
&&
\qquad\qquad\qquad\qquad
-\Big[\frac\la{1-\la}\Box(R^{\mu\a}+P^{\mu\a}) +\Big(\frac\la{1-\la}\Big)^2\nabla^\mu\nabla_\b(R^{\b\a}
 +P^{\b\a})\Big]\,\nabla_\a\nabla_\nu\frac1{F_1^2F_2^2}
\nn
&&
\qquad\qquad\qquad\qquad
+\frac12\frac{\la^2}{1-\la}(\nabla^\mu\nabla_\nu R)\,\Box \frac1{F_1^2F_2^2}+\frac{2\la}{1-\la}
(\nabla_\a\nabla^\mu R_{\nu\b})\nabla^\a\nabla^\b\frac1{F_1^2F_2^2}
\nn
&&
\qquad\qquad
\qquad\qquad
+\frac{\la}{1-\la}\Big[\nabla_\a\nabla_\nu R+\Box R_{\a\nu}\Big]\, \nabla^\mu\nabla_\a\frac1{F_1^2F_2^2}+...\Big\}.
\label{32}
\eeq
Using the results in Appendix~\ref{formula} in (\ref{32}), one gets the result for $M_1$-term:
\begin{eqnarray}
-(KM)^\mu_\nu\frac1{F_1^2F_2^2}
&=& g^{1/2} \frac{1-\la}{16\,\pi^2m^2}\Bigg\{ (\nabla^\mu\nabla_\nu R)\, \left[\, -\frac\la2\,I_{1,0}
+\Big(\frac{\gamma^2}4-\gamma\la -\gamma\Big)\,I_{1,1}\right]
\nn
&&+(\Box P^\mu_\nu)\, \Big[I_{1,0}+\frac{\gamma}{2} I_{1,1}\Big]
+(\nabla^\mu\nabla_\a P^\a_\nu) \Big[ \gamma I_{1,0}+\frac{\gamma^2}{2} I_{1,1}\Big]\Bigg\},
\end{eqnarray}
where the integrals $I_{n_1,n_2}$ here and below are defined in \eqref{36} in Appendix~\ref{formula} and
\begin{eqnarray}
\gamma=\frac\la{1-\la}.
\end{eqnarray}
Note that $\Box R^\mu_\nu$-term does not appear in this order at all.

\subsection{$M_2$-term}

Using (\ref{29}), we have for the third term in \eqref{12}:
\begin{eqnarray}
(KM_2)^\mu_\nu
&=&
K^\mu_\a [\Box,M^\a_\nu ]\,(F_1+F_2)
+ K^\mu_\a \big[\,\Box,[\Box,M^\a_\nu]\,\big]+...
\nn
&=& [\Box,M^\a_\nu ]\,K^\mu_\a \,(F_1+F_2)
+ \big[\,K^\mu_\a,[\Box,M^\a_\nu]\,\big]\,(F_1+F_2)
\nn
&&
+\big[\,\Box,[\Box,M^\a_\nu]\,\big]\,K^\mu_\a+...\, .
\end{eqnarray}
The derivative operator is put to the right, and to the order we need, we find
\begin{eqnarray}
(KM_2)^\mu_\nu &=& [\Box,M^\mu_\nu] F_2(F_1+F_2) + \frac{\la}{1-\la}[\Box,M^\a_\nu]\nabla^\mu \nabla_\a (F_1+F_2)
\nn
&&
+ \big[\,\Box,[\Box,M^\mu_\nu]\,\big] (F_1+2F_2)
+ \frac{\la}{1-\la} \big[\,\Box,[\Box,M^\a_\nu]\,\big]\nabla^\mu\nabla_\a
\nn
&&
+\, \frac{\la}{1-\la}[\nabla^\mu\nabla_\a, [\Box,M^\a_\nu]](F_1+F_2) + \cdots.
\label{57}
\end{eqnarray}

Using the results in Appendix~\ref{formula}, we can calculate each term in \p{57}.
Collecting these terms, we find the $M_2$-term which reads
\beq
&&
(KM_2)^\mu_\nu \frac{1}{F_1^3 F_2^3}
\nn
&&
\qquad
= \,- \frac{g^{1/2}}{16\pi^2 m^2} \bigg\{
\frac{\la}{4} \Big[ 2\la I_{1,1} +\la I_{2,0}
+ (\c-4\la) I_{2,2} + (\c-3\la) I_{1,2}+(\c-\la) I_{2,1}\Big]
\Box R^\mu_\nu
\nn
 &&
\qquad
 + \frac{1}{4} \Big[ 2+4(1-\la) I_{1,0} +4(2-\la) I_{1,1}
+ 2(4-3\la) I_{2,0} + \la(\c+2) I_{1,2}
\nn
&&
\qquad
+ \la(\c+6) I_{2,1}+\c\la I_{2,2} \Big]\Box P^\mu_\nu
\nn
&&
\qquad
 + \frac{\c\la}{8} \big( I_{1,2} + I_{2,1} + I_{2,2} \big)
  \Big[\d^\mu_\nu (2\Box R+\Box P +2 \nabla^\a \nabla^\b P_{\a\b})
 + 2 \nabla^\mu \nabla_\nu P\Big]
\nn
&&
\qquad
+ \frac{\la}{4} \Big[
    (3\c-4\la) (I_{2,1}
 + I_{2,2})
+ (3\c-6\la) I_{1,2}
-2 (1-\la) (I_{2,0} + 2I_{1,1})
\Big] \nabla^\mu\nabla_\nu R
\nn
&&
 \qquad
+ \frac{\la}{4}
\Big[ (\c+8) I_{2,1} +(\c+4) I_{1,2}
+ 2\c I_{2,2} \Big] \nabla_\nu \nabla^\a P^\mu_\a
\nn
&&
 \qquad
+  \frac{\la}{4}\Big[ 8 I_{1,1} + 4 I_{2,0}
+ 3\c I_{1,2} +(3\c+4) I_{2,1} +2\c I_{2,2} \Big]
\nabla^\mu\nabla^\a P_{\a\nu}
\bigg\}.
\label{m2}
\eeq

\subsection{$M_3$-term}

We find that the $M_3$-term is
\begin{eqnarray}
&&
- (KM_3)^\mu_\nu
=  - \Big[4\c \nabla_\c \nabla_\b(R^\mu_\a+P^\mu_\a)
\nabla^\c\nabla^\b\nabla^\a\nabla_\nu F_2
+ 4 (\nabla^\b \nabla^\a P^\mu_\nu)\nabla_\b\nabla_\a F_2^2
\nn
&&
- 4\la (\nabla_\c \nabla_\b R_{\nu\a}) \nabla^\c\nabla^\b\nabla^\mu\nabla^\a F_2
+ 4\c^2 \nabla_\c \nabla_\b
(R^\r_\a+P^\r_\a)\nabla^\mu\nabla_\r\nabla^\c
\nabla^\b\nabla^\a\nabla_\nu
\nn
&&
+4\c (\nabla^\b \nabla^\a P^\r_\nu)\nabla^\mu\nabla_\r\nabla_\b\nabla_\a F_2
- 4\c\la (\nabla_\c \nabla_\b R_{\a\nu}) \Box \nabla^\mu\nabla^\c\nabla^\b\nabla^\a \Big] (F_1+F_2)^2 .
\end{eqnarray}
By use of the results in Appendix~\ref{formula}, this gives
\begin{eqnarray}
&&
- (KM_3)^\mu_\nu \frac{1}{F_1^4 F_2^4}
\nn
&&
 \qquad
= \frac{g^{1/2}}{16\pi^2 m^2} \Bigg[
(\Box R^\mu_\nu) \la \Big(\frac{-3I_{1,2}+3I_{2,1}+I_{3,0}}{6}
+\c \frac{I_{1,3}+3I_{2,2}+I_{3,1}}{6}-\frac{I_{0,3}}{6} \Big)
\nn
&&
\qquad
 + (\Box P^\mu_\nu) \Big( \frac13 + 2(1-\la)(I_{1,1}+I_{2,0}) + \la \frac{3I_{1,2}+6I_{2,1}+I_{3,0}}{3}
+\c \la \frac{I_{1,3}+3I_{2,2}+I_{3,1}}{6} \Big)
\nn
&&
\qquad
 + ( \nabla^\mu \nabla^\a P_{\nu\a}+\nabla_\nu \nabla^\a P^\mu_\a)
\la \Big( \c \frac{I_{1,3}+3I_{2,2}+I_{3,1}}{3}+\frac{3I_{1,2}+6I_{2,1}
+I_{3,0}}{3} \Big)
\nn
&&
\qquad
 + (\nabla^\mu \nabla_\nu R) \la \Big( \frac{-3I_{1,2}+3I_{2,1}
 +I_{3,0}-I_{0,3}}{6}
+\c \frac{I_{1,3}+3I_{2,2}+I_{3,1}}{2} \Big)
\nn
&&
\qquad
 + \c\la \Big( \d^\mu_\nu \Big\{ 2\Box R+\Box P
+2\nabla^\a \nabla^\b P_{\a\b} \Big\}
+2\nabla^\mu \nabla_\nu P \Big) \frac{I_{1,3}+3I_{2,2}+I_{3,1}}{12}
\Bigg] .
\end{eqnarray}

\subsection{The total result}

Collecting all the results up to the $M_3$-terms and substituting the
result of the integrals from Appendix~\ref{formula}, we get
\begin{eqnarray}
&&
- \tr \hat G(x,x)
\sim \frac{g^{1/2}}{360 (4\pi)^2 m^2(1-\la)\la^2}
\Bigg[ \la\Big\{ (60+168 \la-133\la^2)\Box R
\nn
&&
\qquad
\qquad\qquad
\qquad
-10 (6-21\la+8\la^2) (\Box P)
-10(-24+6\la+13\la^2)(\nabla^\a\nabla^\b P_{\a\b}) \Big\}
\nn
&&
\qquad\qquad
\qquad
\qquad
+30(1-\la) \Big\{ (2+5\la-\la^2) (\Box R)
+2(2\la-1) (\Box P)
\nn
&&
\qquad\qquad
\qquad
\qquad
+ 2(4+\la)(\nabla^\a\nabla^\b P_{\a\b})\Big\} \log(1-\la) \Bigg].
\label{re1}
\end{eqnarray}

On the other hand, the result in~\cite{Gusynin99} is
\begin{eqnarray}
\tr \hat E_4(x)
\,=\,
\frac{1}{(4\pi)^2}\,a^\mu_{2, \mu}(x,x)
\,=\, \frac{1}{(4\pi)^2}
\big(c_1 \Box P+ c_2 \nabla^\a\nabla^\b P_{\a\b}+ c_{11}\Box R\big),
\end{eqnarray}
with the coefficients given in \eqref{GuKo-E4}.
Thus, this result agrees with Eq.~(\ref{re1}).

\section{Final result for the anomalous $\Box R$ term}
\label{sec5}

Let us now apply the algorithm for the minimal fourth-order
operator  (\ref{Gusyn90}) and for the nonminimal vector
operator (\ref{nonmin-GK}), 
to derive the one-loop total-derivative divergence in the theory
(\ref{action conf}).

We start from the tensor sector. The intermediate expressions
for the elements of the general formula (\ref{Gusyn90}), with the
elements (\ref{Vrhosi}) -- (\ref{U}), can be
found in Appendix~\ref{secApB}. Summing up these expressions,
we arrive at
\bea
&&
{\rm tr}\,\hat a_2^H
\,=\,
\frac12\, \tr
\bigg\{ \frac{74}{135} \d_{\mu\nu,\a\b}\Box R
+ \frac{8}{27} g_{\mu\nu} \Box R_{\a\b}
- \frac83  g_{\nu\b}\Box R_{\mu\a}+\frac23\,\Box R_{\mu\a\nu\b}
\qquad\quad
\nn
&&
\qquad\qquad
\qquad
- \frac{20}{27}\,\nabla_\mu\nabla_\nu R_{\a\b}
+\frac49\,\nabla_\mu\nabla_\a R_{\nu\b}
+\frac{20}{27}\,g_{\nu\b} \nabla_\mu\nabla_\a R \bigg\}.
\label{eq8}
\eea

The rule of taking trace should take into account that the unite
matrix is the projector to the traceless states, such that, e.g.,
\beq
\tr \,\delta_{\mu\nu,\a\b}
\,=\,\Big(\d^{\mu\nu,\a\b} - \frac14 g^{\mu\nu} g^{\a\b}\Big)
\d_{\mu\nu,\a\b} \,=\, 9.
\eeq
In this way, after small algebra, we get
\beq
{\rm tr}\,\hat a_2^H=\frac{13}{135}\,\Box R.
\label{eq9}
\eeq

For the nonminimal vector operators, we have to apply the
results (\ref{nonmin-GK}) and (\ref{GuKo-E4}) to the
operators (\ref{Y}) and (\ref{FPop}). In the first case, $P^{\mu\nu}=0$ and $\la = 1/3$. After a small algebra, we get
\beq
{\rm tr}\,\hat a_2^Y=
\bigg\{
\frac{911}{720}\,-\, \frac{8}{3}\,\log\Big(\frac32\Big)
\bigg\}\Box R.
\label{trlogY}
\eeq
For the ghost operator $\hat{M}$, we have $P^{\mu\nu}=R^{\mu\nu}$
and $\la = -1/2$. The calculations give, in this case,
\beq
{\rm tr}\,\hat a_2^M=
\bigg\{ \frac{247}{540}\,-\,\frac{5}{12}\,\log\Big(\frac32\Big)
\bigg\}\Box R.
\label{trlogM}
\eeq

Substituting these results in (\ref{FHanomaly2}) we get
\beq
\langle\, T^{\mu}_\mu(x)\rangle\,\big|_{\,total\;\;derivative}=
\frac1{16\pi^2}\bigg\{ \frac{7}{2}\,\log \Big(\frac32\Big)
\,- \, \frac{159}{80}
\bigg\}\Box R,
\label{final}
\eeq
where we explicitly indicated that this is a total derivative part
of the full trace anomaly. Other terms in the anomaly can be recovered from the integrand of the one-loop divergences in Weyl gravity model, which can be found in~\cite{antmot} or~\cite{Weyl,OP2015}.
As we have mentioned above, the result quoted in  (\ref{final})
cannot be modified by adding a finite classical $R^2$ term, and the last
enters into the anomaly-induced action only in the form defined by the
anomaly and the relation (\ref{identity-a}).

At this point, we can formulate the complete version of
Eq.~(\ref{anomaly}) for the effective action of gravity induced
by conformal anomaly in Weyl-squared quantum conformal gravity.
This equation has the form
\beq
\langle\, T^{\mu}_\mu(x)\rangle=-\frac{2}{\sqrt{g}}
\,g_{\mu\nu}\frac{\de{\Ga}_{renorm}}{\de g_{\mu\nu}}
\,=\,  \om C^2 + b E_4 + c \Box R ,
\label{TinW2}
\eeq
where we stressed the fact that the equation is for the finite renormalized part $\Gamma_{renorm}$
of the effective action with the divergences are subtracted in the process of renormalization.
According to~\cite{antmot,Weyl,OP2015} and (\ref{final}) the coefficients are
\beq
&&
\om \,=\, \frac{1}{(4\pi)^2}\,\frac{199}{15}\,,
\nonumber
\\
&&
b \,=\, -\,\frac{1}{(4\pi)^2}\,\frac{87}{20}\,,
\nonumber
\\
&&
c \,=\, \frac{1}{(4\pi)^2}\bigg[
\frac{7}{2}\,\log \Big(\frac32\Big) \,- \, \frac{159}{80}
\bigg].
\label{ombc}
\eeq
The solution of Eq.~(\ref{TinW2}) has the general form
(\ref{abc-terms}) and, as usual, it includes a conformal
invariant functional of the metric $S_c$, playing the role of
an integration constant for the equation. This functional is not
controlled by the trace anomaly.

\section{Conclusions}
\label{conclusions}

We have presented original calculation of the $\Box R$ term in the one-loop
trace anomaly of Weyl gravity model with the action
(\ref{action conf}). The mapping of this anomaly to one-loop divergences of the theory runs within
the framework of the zeta-function regularization or the regularization by the covariant proper time
cutoff in the heat kernel. Functional integration of the $\Box R$ anomaly term yields a finite local
$R^2$ term in the one-loop effective action. Unlike the semiclassical theory, this term cannot
be ``removed'' by adding a ``finite counterterm''. This situation
demonstrates that the local conformal invariance gets violated at
the one-loop level. This violation is related not only to the
nonlocal terms shown in (\ref{abc-terms}), but also to the local
$R^2$ term. Starting from the second loop, one has to take this
term into account, modify propagator by adding the dynamical
scalar mode, modify vertices, etc. This confirms that the local
conformal symmetry cannot be exact at the quantum level and,
moreover, its violation beyond one-loop approximation cannot
be controlled~\cite{Capper1975} unless some mechanisms like supersymmetry are used for the cancellation
of anomalies~\cite{FrTs84}.

Despite this set of intrinsic inconsistencies and regularization ambiguities of the above type,
the calculation of anomalous $\Box R$ terms still makes sense for the sake of the potential analysis of
the boundary terms in one-loop divergences and in view of cosmological implications of the related $R^2$ terms
in the Starobinsky model of modified gravity theory. Let us also note that since in dimensional and
Pauli-Villars regularizations there are ambiguities on the way from divergences
to the trace anomaly, it would be certainly interesting to make derivation within nonlocal covariant curvature
expansion of~\cite{CPTII,CPTIII}. In particular, it is worth deriving the form factors of the $R^2$ term
for the minimal fourth-order and nonminimal second-order vector operators. However, this challenging
calculation is beyond the scope of the present work and we postpone it for future.

Let us mention two important aspects of the conformal anomaly.
First, while the local conformal symmetry is violated by both
local and non-local terms, the \textit{global} symmetry still holds
in the anomaly-induced action (\ref{abc-terms}). This shows that
the destinies of these two symmetries at the quantum level are very
much different. Unlike the anomaly which was discussed
above, the violation of the global symmetry requires the presence
of a dimensional parameter, which may emerge either from an
interaction with massive fields or from the phase transition and
dimensional transmutation, as discussed in~\cite{Buch87, induce}.

The only way to use local conformal symmetry in quantum theory
is by assuming the corresponding hierarchy, as it was discussed in~\cite{Weyl}.
One can start with the theory that has both $C^2$ and
$R^2$ terms, but the coefficient of the last is very small, such that
its contributions in loops get strongly suppressed. There is a chance
that this hierarchy may hold at higher loops. This scheme enables
one to preserve the advantages of conformal theory, including the
compact and useful form of anomaly-induced action.

\section*{Acknowledgments}

The work of A.O.B. was supported by the Russian Science Foundation grant No 23-12-00051.
The work of N.O. was supported in part by the Grant-in-Aid for
Scientific Research Fund of the JSPS (C) No. 20K03980, and by
the Ministry of Science and Technology,
R. O. C. (Taiwan) under the grant MOST 112-2811-M-008-016.
The work of I.Sh. was partially supported by Conselho Nacional de
Desenvolvimento Cient\'{i}fico e Tecnol\'{o}gico - CNPq under the
grant 303635/2018-5.

\appendix


\section{Details of the calculatons}
\label{formula}

In this appendix, we summarize formulas necessary for the evaluation of the
Green's functions performed in Sec.~\ref{s-nonmi}.

$M_2$ and $M_3$ take the following skeleton form in the
needed approximation
\begin{eqnarray}
&&M_2=[\Box,M]\,(F_1+F_2) +\big[\,\Box,[\Box,M]\,\big]+...,
\label{29}\\
&&M_3=\big[\,\Box,[\Box,M]\,\big]\,(F_1+F_2)^2+...,
\label{30}
\end{eqnarray}
and the expansion (\ref{12}) starts with the following four terms
which contribute to needed $\nabla\nabla R$
and $\nabla\nabla P$ structures
\begin{eqnarray}
\frac1{F-m^2}=K\frac1{F_1F_2}-KM\frac1{F_1^2F_2^2} +KM_2\frac1{F_1^3F_2^3}-KM_3\frac1{F_1^4F_2^4}+...\,,
\label{11a}
\end{eqnarray}

Note that the coefficient functions in (\ref{11a}) are differential
operators $K$, $KM$, $KM_2$, $KM_3$ which contain the mass
parameter only in combinations $F_1,F_2$ given by (\ref{11}) or
their powers. Therefore the final answer is a linear combination
of the following massive universal traces
\beq
&&
\nabla_{\a_1}...\nabla_{\a_{2n}} \frac1{F_1^{n_1} F_2^{n_2}}
\delta(x,y)\,\Big|_{y=x},
\eeq
where $\quad n_1+n_2-n=3$ and the restriction on $n$, $n_1$ and $n_2$
follows from their dimensionality $\sim 1/m^2$.
To evaluate this, we need the proper time representation in terms of the heat kernel
\beq
&&
\frac1{(\Box-m^2_1)^{n_1}\, (\Box-m^2_2)^{n_2}}\delta(x,y)
\nn
&&
\qquad
= \frac{(-1)^{n_1+n_2}}{\varGamma(n_1)\varGamma(n_2)}
\int_0^\infty ds_1\,ds_2\,s_1^{n_1-1}\,
s_2^{n_2-1}  e^{(s_1+s_2)\Box-(s_1m_1^2+s_2m_2^2)}\delta(x,y)
\nn
&&
\qquad
= \frac{(-1)^{n_1+n_2}}{\varGamma(n_1)\varGamma(n_2)}
\int_0^\infty dt\,t^{n_1+n_2-1}
\int_0^1 d\a\,\a^{n_1-1}(1-\a)^{n_2-1}\,e^{-tm^2(\a)}
\,e^{t\Box}\delta(x,y),\quad
\label{19}
\eeq
where we have made the change of variables
$s_1=\a t, s_2=(1-\a)t$ and $m^2(\a)\equiv m_1^2\a+m_2^2(1-\a)$.
Substitution of the Schwinger-DeWitt expansion for the
$\Box$-operator~\cite{DeWitt65}
\beq
&&
e^{s\Box}\delta^\mu_\nu(x,x')
\,=\,
\frac{\Delta^{1/2}(x,x')}{(4\pi s)^{d/2}}\,g^{1/2}(x')
\,e^{-\frac{\sigma(x,x')}{2s}}\,
 \Big(\,a^{\Box\,\mu}_{0\,\nu}(x,x')
 \nn
 &&
 \qquad
 \qquad
 \qquad
 + \,\,
 a^{\Box\,\mu}_{1\,\nu}(x,x')\,s +a^{\Box\,\mu}_{2\,\nu}(x,x')\,s^2+...\Big),~~
\label{Sch-Dew2}
\eeq
then gives the final result with a given accuracy in curvatures.
Here $\sigma(x,x')$ is a geodesic interval given by one half of the square of the distance along the geodesic
between $x$ and $x'$, $\Delta(x,x')= - g(x)^{-1/2}\det(-\s_{,\,\mu\nu'}) g(x')^{-1/2}$,
and $a^{\Box\,\mu}_{n\,\nu}(x,x')$ are the Schwinger-DeWitt coefficients for the $\Box$-operator.

We need to keep only the first term to obtain
\begin{eqnarray}
&& g^{-1/2}\nabla_{\a_1}...\nabla_{\a_{2n}} \frac1{F_1^{n_1} F_2^{n_2}}\delta(x,y)\,\Big|_{y=x}
\nn
&&
\qquad
=\frac{(-1)^{n_1+n_2}}{\varGamma(n_1)\varGamma(n_2)}\int\limits_0^\infty \frac{dt\,t^{n_1+n_2-1}}{(4\pi t)^2}
\int_0^1 d\a\,\a^{n_1-1}(1-\a)^{n_2-1}\,
\nn
&&
\qquad\qquad
\times
\quad
e^{-tm^2(\a)}\,\nabla_{\a_1}...\nabla_{\a_{2n}}
e^{-\sigma(x,y)/2t}\,\Big|_{y=x}+...
\nn
&&\qquad= -\frac1{2^n(n_1-1)!\,(n_2-1)!}\,\frac{1-\la}{16\,\pi^2m^2}
 \int\limits_0^1 d\a\,\frac{\a^{n_1-1}(1-\a)^{n_2-1}}{1-\a\la} \,g^{(n)}_{\a_1...\a_{2n}}+...\,,
\label{35}
\end{eqnarray}
where we have taken into account that $(-1)^{n+n_1+n_2}=-1$ in view
of the above restriction, ellipses denote terms other than $(\Box R)$
and $(\Box P)$-structures, and we have used the totally symmetric
tensor built of the metric~\cite{bavi85}
\begin{eqnarray}
&&\nabla_{\a_1}...\nabla_{\a_{2n}} e^{-\sigma(x,y)/2t}\,\Big|_{\,y=x}+...
=(\nabla_{\a_1}-\sigma_{\a_1}/2t)...
 (\nabla_{\a_{2n}} -\sigma_{\a_{2n}}/2t)\,1\,\Big|_{\,y=x}+...
\nn
&&\qquad\qquad\qquad\qquad\qquad\qquad=\left(-\frac1{2t}\right)^n \,g^{(n)}_{\a_1...\a_{2n}}+...\,,\\
&&g^{(1)}_{\a\b}=g_{\a\b},\quad g^{(2)}_{\a\b\mu\nu}=g_{\a\b}g_{\mu\nu}+g_{\a\mu}g_{\b\nu}
 +g_{\a\nu}g_{\b\mu},\quad ...
\end{eqnarray}

It proves useful to introduce the notation for the $\a$-integrals,
\begin{eqnarray}
 I_{n_1,n_2}(\la)=\int\limits_0^1 d\a\, \frac{\a^{n_1}(1-\a)^{n_2}}{1-\a\la}.
\label{36}
\end{eqnarray}
This integral can be easily evaluated for integer $n_1, n_2$ by
using Mathematica or other software.

Next, it follows from (\ref{35}) that for all $n_1$, $n_2$ and the
number of derivatives $2n=2(n_1+n_2-3)$, the needed massive
universal functional traces read
\begin{eqnarray}
\nabla_{\a_1}...\nabla_{\a_{2n}} \frac1{F_1^{n_1} F_2^{n_2}}
\delta(x,y)\,\Big|_{\,y=x}
&=&
g^{1/2} \frac{1-\la}{16\,\pi^2m^2} \,\frac{(-1)^{n+n_1+n_2}
\,I_{n_1-1,n_2-1}(\la)}{2^n\, (n_1-1)!\,(n_2-1)!}
 \,g^{(n)}_{\a_1...\a_{2n}}+...\,
\nn
\hs{-2}&=&\hs{-2}
- g^{1/2} \frac{1-\la}{16\,\pi^2m^2}
\,\frac{I_{n_1-1,n_2-1}(\la)}{2^n\, (n_1-1)!\,(n_2-1)!}
\,g^{(n)}_{\a_1...\a_{2n}}+... \hs{3}
\end{eqnarray}

We also need
\begin{eqnarray}
\nabla_{\a_1}...\nabla_{\a_{2n}} \frac1{F_1^{n_1}}\delta(x,y)\,\Big|_{\,y=x} .
\end{eqnarray}
First recall that
\begin{eqnarray}
\frac{1}{F_1^{n_1}} = \frac{(-1)^{n_1}}{\G(n_1)} \int_0^\infty ds s^{n_1-1}e^{sF_1}.
\end{eqnarray}
So we get
\begin{eqnarray}
\nabla_{\a_1}...\nabla_{\a_{2n}}\frac{1}{F_1^{n_1}}\delta(x,y)\,\Big|_{\,y=x}
&=& g^{1/2} \frac{(-1)^{n_1}}{\G(n_1)} \int_0^\infty ds s^{n_1-1}\nabla_{\a_1}...\nabla_{\a_{2n}}
e^{sF_1} \delta(x,y)\,\Big|_{\,y=x}
\nn
&=& g^{1/2} \frac{(-1)^{n_1}}{\G(n_1)} \int_0^\infty \frac{ds s^{n_1-1}}{(4\pi s)^2}
\nabla_{\a_1}...\nabla_{\a_{2n}} e^{-sm^2}
e^{-\sigma(x,y)/2s}\Big|_{\,y=x}
\nn
&=& g^{1/2} \frac{(-1)^{n_1}}{16\pi^2 \G(n_1)}\Big(-\frac{1}{2}\Big)^n \int_0^\infty ds e^{-sm^2}
g^{(n)}_{\a_1...\a_{2n}} +...
\nn
&=& -\frac{g^{1/2} }{16\pi^2 m^2} \frac{1}{2^n (n_1-1)!} g^{(n)}_{\a_1...\a_{2n}} + ...\, ,
\end{eqnarray}
where we have chosen $n=n_1-3$.

\section{Tensor contribution in $\Box R$-type divergences}
\label{secApB}

Let us list the particular results necessary to evaluate the general formula (\ref{Gusyn90}) in (\ref{eq8})
for the tensor sector. These are derived from (\ref{Vrhosi}) and (\ref{Nrho}).
\beq
\big(\nabla_\rho\nabla_\si\hat{V}^{\rho\si}\big)_{\mu\nu,\a\b}
&=&
\frac13\d_{\mu\nu,\a\b} \pare{\Box R}
-\frac83 \pare{\nabla_\mu\nabla_\a R} g_{\nu\b}
+\frac43 \pare{\nabla_\mu\nabla_\nu R_{\a\b}}
+ \frac43 \pare{\nabla_\a\nabla_\b R_{\mu\nu}}
\nn
&&
+ 2 \pare{\nabla_\nu\nabla_\b R_{\mu\a}}
+ 4 \pare{\Box R_{\mu\a\nu\b}},
\label{eq5}
\\
\big(\Box \hat{V}\big)_{\mu\nu,\a\b}
&=&
\frac23\d_{\mu\nu,\a\b}\pare{\Box R}
+\frac43g_{\mu\nu}\pare{\Box R_{\a\b}}
+\frac43g_{\a\b}\pare{\Box R_{\mu\nu}}
\nonumber
\\
&&
- 6\pare{\Box R_{\mu\a}}g_{\nu\b}
+ 16 \pare{\Box R_{\mu\a\nu\b}},\qquad
\label{eq6}
\\
\big(\nabla_\lambda\hat{N}^\lambda\big)_{\mu\nu,\a\b}
&=&
\frac13\d_{\mu\nu,\a\b} \pare{\Box R}
-\frac43 \pare{\nabla_\mu\nabla_\nu R_{\a\b}}
-\frac83 \pare{\nabla_\mu\nabla_\a R} g_{\nu\b}
-2 \pare{\nabla_\mu\nabla_\a R_{\nu\b}}
\nn
&&
+4 \pare{\nabla_\a\nabla_\b R_{\mu\nu}}
+4 \pare{\nabla_\nu\nabla_\b R_{\mu\a}}
+4 \pare{\Box R_{\mu\a\nu\b}},
\label{eq7}
\eea
where symmetrization $\mu\leftrightarrow \nu$, $\a\leftrightarrow \b$ and $(\mu,\nu)\leftrightarrow (\a,\b)$
should be understood.


\bibliographystyle{unsrturl}

\end{document}